\documentclass[twocolumn,amsmath,showkeys,amssymb,preprintnumbers,nofootinbib]{revtex4}

\usepackage{graphicx}
\usepackage{dcolumn}
\usepackage{bm}

\begin{document}

\preprint{To appear in {\it Int. J. Modeling Simulation and Scientific Computing}}

\markboth{Authors' Names}{Paper's Title}

\title{Geant4 Applications for Modeling Molecular Transport \\in Complex Vacuum Geometries}

\author{JACK SINGAL}
\affiliation{Department of Physics, University of Richmond \\ Gottwald Center for the Sciences \\ 28 Westhampton Drive, Richmond, VA 23173 \\ jsingal@richmond.edu}

\author{J. BRIAN LANGTON, RAFE SCHINDLER}
\affiliation{Kavli Institute for Particle Astrophysics and Cosmology \\ SLAC National Accelerator Laboratory and Stanford University\\ Menlo Park, CA 94025}

\begin{abstract}
We discuss a novel use of the Geant4 simulation toolkit to model molecular transport in a vacuum environment, in the molecular flow regime.  The Geant4 toolkit was originally developed by the high energy physics community to simulate the interactions of elementary particles within complex detector systems.  Here its capabilities are utilized to model molecular vacuum transport in geometries where other techniques are impractical.  The techniques are verified with an application representing a simple vacuum geometry that has been studied previously both analytically and by basic Monte Carlo simulation.  We discuss the use of an application with a very complicated geometry, that of the Large Synoptic Survey Telescope camera cryostat, to determine probabilities of transport of contaminant molecules to optical surfaces where control of contamination is crucial. 
\end{abstract}


\maketitle

\section{Introduction} \label{intro}

The cryostat of the Large Synoptic Survey Telescope camera (LSSTCam)\cite{LSSTover,SPIE1} represents a significant challenge for vacuum design.\cite{RSI}  With a volume of 2.9 m$^3$, it will contain nearly 1000 kg of material, and have more than 40 m$^2$ of exposed surface area within.   The geometry of the LSSTCam cryostat presents an environment too complicated for vacuum transport to be modeled by analytical calculations, or by techniques such as thermal conductance calculations in conjunction with finite element analysis.\cite{AnsysP}  Instead, a fully three dimensional model of the geometry is required, with a full Monte Carlo treatment of the initial conditions and subsequent propagation of particles through their interactions with surfaces.  Monte Carlo calculations of vacuum transport have been performed extensively via stand-alone custom code, but for relatively simpler geometries.\cite{Davis,Vaktrak}  The complexity of the LSSTCam cryostat would render the specification of the geometry by similar stand-alone code impossible.  Some complex vacuum geometries have be modeled in the MolFlow+ simulation suite.\cite{Molflow}  However, for even more complicated geometries such as in the LSSTCam cryostat, the Geant4 simulation toolkit\cite{Geant1,Geant2} presents itself as an ideal platform for developing an application to model vacuum transport.

\begin{figure}
\includegraphics[width=3.0in]{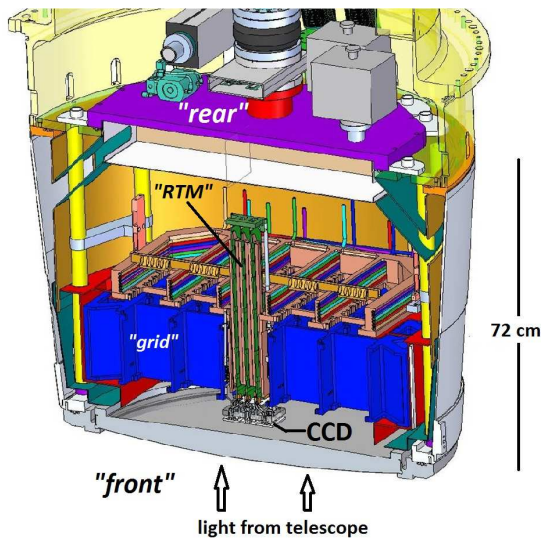}
\caption{CAD rendering of an LSST camera cryostat design in cross section.  A brief summary of the cryostat is given in \S \ref{intro} and its geometry is discussed in \S \ref{setup}.  Shown here is one science raft tower module (RTM) containing CCD detectors and electronics boards and other structures, the grid (rendered in blue), the cryoplate (pink and dark orange), the cold shroud (red), the lower shroud (orange), some of the pumping plenum and chimneys (dark green and white), the feed-thru flange (transparent light green), and the pump plate (blue). The cryostat body and the L3 lens are rendered in grey.  This particular scenario is known as ``option 3'', however the differences between the various scenarios are subtle enough not to dominate a large-scale view.  Some structures are not shown. }
\label{cryo}
\end{figure}

\begin{table}[ph]
\caption{ {\scriptsize Test case described in \S \ref{confirm} --- probability of a particle first reaching the opposite end rather than the originating end of a pipe of length $L$ and radius $R$ via molecular vacuum transport.  $P_e$ is the probability as determined by analytical calculations and reported in Clausing\cite{Clausing}, $P_m$ is the probability as determined from Monte Carlo methods by Davis\cite{Davis}, and $P_G$ is the probability as determined by a Geant4 application described in \S \ref{confirm} with 10000 randomized runs. }}
\begin{tabular}{cccc} 
 \toprule
$L$/$R$  &  $P_e$  &  $P_m$  &  $P_G$  \\  \colrule
 0.5  &  0.80  & 0.80 &  0.80  \\
 1.0  &  0.67  & 0.68 &  0.67  \\ 
 1.5  &  0.58  & 0.57 &  0.57  \\ 
 2.0  &  0.51  & 0.53 &  0.51  \\ 
 3.0  &  0.42  & 0.43 &  0.44  \\ 
 4.0  &  0.36  & 0.36 &  0.35  \\ 
 5.0  &  0.32  & 0.32 &  0.29  \\ 
 10.0  &  0.20  & --- &  0.18  \\
 20.0  &  0.11  & --- &  0.10  \\ \colrule
\end{tabular} 
\label{resu}
\end{table}

Geant4 is a Monte Carlo simulation toolkit used extensively in particle physics, nuclear physics, medical physics, and aerospace applications.  It is a flexible, self-contained software package that allows the definition of a volume containing real materials arranged into complex geometries and then the transport of elementary or composite particles within the volume.  Geant4 follows any fundamental interactions of the particles with the materials, as well as any products of those interactions.  The toolkit consists of a large number of C++ classes which the application developer uses in order to create a specific application that models a physical system. 

In this letter we describe Geant4 applications that were developed to model vacuum transport in the molecular flow regime.  The general strategy is summarized in \S \ref{ao}.   In \S \ref{confirm} an application with a simple geometry to verify the techniques used is discussed.   In \S \ref{setup} a specific application to the LSSTCam cryostat is described.  Several cases of the cryostat geometry and initial conditions of the molecular distribution are evaluated to demonstrate the power and flexibility of this design tool.   These general techniques could of course be applied to any vacuum geometry, and this is explored in \S \ref{disc}.\footnote{All applications discussed in this work are available at http://sourceforge.net/projects/geant4vacmod/.}

\section{Application Strategy} \label{ao}

To develop a Geant4 application, particles, their interactions, a detector geometry, and relevant cataloging of outcomes (known as ``scoring'') must be implemented.  In Geant4 particles are specified by properties including mass, and charge, and, reflecting its origin in the particle physics community, quantities such as spin, quark contents, and lifetime and decay modes.  There are many pre-defined particles (e.g. electrons, positrons, photons, etc), and application developers can create their own by specifying the relevant properties.  Particles can interact with other particles and surfaces, as well as leave the area of interest or decay into other particles if possible, as specified by the application developer.  The possible interaction outcomes and their probabilities for any particle type with any other particle type or surface can be specified by implementations of included classes.  In Geant4, evolution of a system moves the particles through tiny steps, at the end of which it is evaluated whether the particle has encountered any other particles or any surfaces, and if so, an interaction is carried out.  Any original and subsequently created particles are tracked until they reach a defined outcome, such as encountering a given surface or exiting the detector, or when a defined amount of time has elapsed.  Scoring involves counting the number of times a certain event has happened, for example a type of particle decaying or touching a given region of a surface.  Recent years have seen applications using the capabilities of Geant4 extend beyond the original fields, such as in modeling phonon propagation in a silicon crystal for dark matter direct detection\cite{CDMS} and modeling the damage to DNA from radiation in various environments\cite{Geant4DNA}.  

To model molecular transport in vacuum with the Geant4 toolkit, the vacuum space of interest is defined as the detector geometry.  All structures are specified according to their material (stainless steel, copper, G10, silicon, and so on) in the standard Geant4 way taking advantage of the Geant4 material database, which uses National Institutes of Standards and Technology (NIST) standard values.  They are also assigned their predicted operating temperatures. Custom particles are specified with implementations of G4ParticleDefinition, to represent the desired molecular species (``Waterons'' and ``Nitroginos'' for example, representing water and nitrogen molecules, respectively) that lack internal structure and charge but have the proper molecular weight, allowing, for example, the mean velocities and residence times to be affected by contact with surfaces of different temperatures.  

There are no particle-particle interactions, as the physical situations being modeled are high vacuum systems where transport is in the molecular flow regime, so that the ``ballistic'' approximation is valid.  For example, the LSSTCam cryostat will have pressures of $\sim$10$^{-6}$ torr upon pump-down and around two orders of magnitude lower when surfaces are cooled to their operating temperatures, so that the molecular mean free path is significantly larger than the dimensions involved.  Particles thus only interact with material surfaces, and then in only a particular way that represents how surfaces direct particle transport in the molecular flow regime.  

When a particle encounters a surface in the simulation, it is adsorbed at the point of contact and then desorbed with the probability of a given direction proportional to the cosine of the angle with the normal vector to the surface at the point.  We specify these interactions in an implementation of G4VDiscreteProcess.  This model of surface interaction is used in other Monte Carlo implementations of molecular flow vacuum transport.\cite{Davis,Vaktrak,Molflow}  It is based on the considerations that the probability of a particle successfully desorbing from a surface onto which it is adsorbed is proportional to the normal component of its velocity away from the surface, which is the cosine of the angle of the velocity vector to the surface normal.  Thus the angular distribution of desorbed molecules is described by a cosine function from the surface normal.  This is discussed from a theoretical perspective in surface chemistry texts\cite{PMC}, and some experimental verifications have been carried out.\cite{Rettner89,Steinruck84}  Like many assumptions in simulations, the cosine to the surface normal desorption probability distribution is a model which represents real-world data in a variety of circumstances, but may not hold on a microscopic level in every single case.\cite{Steinruck84}  However, we believe that following this approach, which has a sound theoretical basis and is the one used in the simulation literature, is the most appropriate.  

An implementation of G4GeneralParticleSource is used to originate particles with initial momentum directions and positions chosen randomly according to distributions appropriate to the application, and from the regions of interest within the volume of the vacuum.  Particles are tagged with the name of the first surface they encounter, which can be considered the originating item, and this is done in an implementation of G4UserSteppingAction with the originating surface name and any other desired information attached with an implementation of G4UserTrackInformation.  

Scoring is implemented such that particles propagate until they reach one of several particular specified terminating outcomes, for example exiting through a pump port, or contacting a particular surface.  These terminating outcomes are implemented through the use of regions specified in the detector construction, which are checked at each step in our implementation of G4UserSteppingAction.  Each molecule's originating item and terminating outcome is cataloged in the output data file, which consists of columns representing the different outcomes with a ``1'' entered into the proper outcome column for each molecule.  In this manner the total number of molecules resulting in each outcome can be easily summed.

\begin{figure}
\includegraphics[width=3.0in]{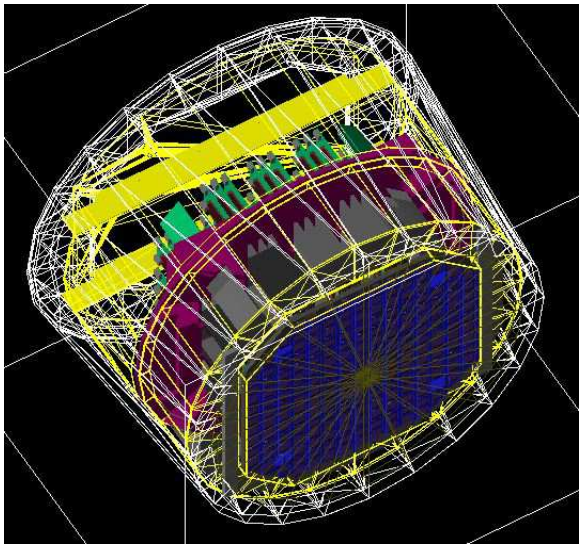}
\caption{ {\scriptsize View of an OpenGL rendering of the geometry used in the LSST camera cryostat Geant4 application.  For clarity, certain structures are rendered in wireframe and others are rendered invisible.  This particular geometrical scenario shown is known as ``option 3'', however the differences between the various scenarios are subtle enough not to dominate such a large-scale view. }}
\label{geo1}
\end{figure}

\begin{table}[ph]
\caption{ {\scriptsize Likelihood of three different outcomes (exiting via a pump port, encountering a getter pump, or encountering the front surface of a CCD) occurring first for molecules in the LSSTCam cryostat application, for several different geometries.  The geometries are discussed in \S \ref{setup}.  Results are shown both for molecules originating with random positions and momentum directions from the entire region within the cryostat behind the focal plane (``All loc.''), and for molecules originating with random positions and momentum directions only within the envelope containing all of the electronics boards (``Elec. reg.'').  Values may not total 100\% due to rounding. }}
\begin{tabular}{cccc} 
 \toprule
{\it All loc.}  &  &   &   \\ 
Geometry &   Pump   &   Getter   & CCD Front Surf.  \\ \colrule
Baseline &  12\%  &  32\% &  56\%  \\
Option 2 &  15\%  &  32\% &  54\%  \\
Option 3 &  8\%  &  36\% &  56\%  \\ 
Option 3 + 2xI &  23\%  &  29\% & 47\%  \\ 
Option 3 + 4xI &  50\%  &  20\% & 30\%  \\ \colrule
{\it Elec. reg.} &  &   &   \\ 
Geometry & Pump & Getter & CCD Front Surf.  \\ \colrule
Baseline &  13\% & 31\%  &  57\%  \\
Option 2 &  13\% & 35\%  &  52\%  \\
Option 3 &  8\% & 37\%  &  55\%  \\ 
Option 3 + 2xI &  24\%  & 24\% &  52\%  \\ 
Option 3 + 4xI &  50\%  & 17\% &  33\%  \\ \colrule
\end{tabular} 
\label{contribs}
\end{table}

\section{Confirmation of techniques with a simple geometry} \label{confirm}

To test the validity of using the Geant4 toolkit to model molecular transport in vacuum with the techniques discussed here, we have developed a Geant4 application with a very simple geometry for which transport results can be compared to those in the literature.  Following Davis \cite{Davis}, the geometry modeled is a right circular cylinder (i.e. a tube), where molecules enter at one end with randomly distributed initial radial and azimuthal positions in the plane of the opening and momentum vectors randomly distributed proportional to the cosine of their angle with the surface normal of the opening plane.  Molecules are propagated through the cylinder with wall interactions as discussed in \S \ref{ao}.  The quantity of interest is whether a molecule first reaches the opposite end of the cylinder or returns to the opening through which it entered.  Thus this is just the problem of molecular transport probability through a pipe of a given length and radius.

The probabilities vary with the ratio of the length ($L$) to the radius ($R$) of the cylinder, and have been calculated analytically by Clausing\cite{Clausing} in 1930 and subsequently by Davis\cite{Davis} using simple Monte Carlo methods.  The Davis code tracks a particle and, if it is coincident with the coordinates of a surface, resets the trajectory to be in another direction with a likelihood probabilty proportional to the cosine of the angle with the normal vector to the surface at the point in question.  A Geant4 application for this geometry is implemented and the results for various $L/R$ values compared.  The application has regions specified at the two ends of the cylinder that terminate a given molecule's trajectory when they are entered, and the terminating region is recorded in the output file as described in \S \ref{ao}.  Table 1 shows the probabilities for a molecule to reach the opposite opening for different values of $L$/$R$, as calculated analytically by Clausing, via Monte Carlo methods by Davis, and by the Geant4 application.  The statistical uncertainty on the probabilities we have obtained can be estimated from standard theory as $\sigma=[\sqrt{n}/M]/M$, for a situation where there are $n$ randomized trials with $M$ possible discrete outcomes.  In this case, $\sigma$=0.005.  The results from the Geant4 application show very good agreement with the probabilities obtained by the other methods.

\section{LSSTCam Cryostat application} \label{setup}

Figure \ref{cryo} shows a CAD rendering of an LSSTCam cryostat layout.  The strict requirements on LSST's photon throughput necessitate a detailed knowledge of the effect of potential molecular contaminants that may deposit on the cold CCD focal plane surface within the cryostat.  One crucial input to any model that addresses this question is the transport of molecules within the cryostat.  In particular, how likely is a contaminant molecule from the bulk of the vacuum space where most of the electronics and other structures are located to reach the focal plane region rather than being removed from the system elsewhere?

The LSSTCam cryostat geometry consists of a significant number of internal structures contained within the cryostat body, which forms the vacuum enclosure.  The body is a truncated conical section with small interior dimension (ID) 0.95 m and large ID 1.06 m.  It is sealed at the smaller ``front'' end (through which light from the telescope enters) by a glass lens designated ``L3''.  Directly behind L3 is the focal plane consisting of the 3.2 billion 10 $\mu$m CCD pixels arranged in 21 raft tower modules (RTMs).  Each RTM contains nine 4kx4k pixel CCD detectors, and is itself a self-contained camera with the CCD packages, a Silicon Carbide ``raft'' on which they are mounted and aligned, conductance barriers, electronics boards, thermal and wall structures, and connectors and cabling.  Four triangular shaped RTMs carrying guide and wavefront sensors are located at the corners of the focal plane.  The RTMs pass through a Silicon Carbide ceramic honeycomb structure known as the grid, which kinematically supports the raft.  Behind the grid, the cryoplate is a mechanical structure that carries the balance of the load of the RTM.  The cryoplate contains cryogenic refrigerant channels that provide cooling to the RTM electronics and other structures.  Behind the cryoplate, a separate cold circuit removes heat from the rearmost electronics at a higher temperature.   There are also various shrouds, chimneys, and plenums that provide thermal radiation shields and direct molecular transport.  The rear of the cryostat is an annular feed-through flange containing hermetic signal and cryogenic feed-throughs, and an octagonal plate containing ports for turbomolecular and ion pumps and gauges.  In combination, they seal the rear of the vacuum space. Two large molecular sieve getter pump structures reside just behind L3, facing the focal plane.

\begin{figure}
\includegraphics[width=3.0in]{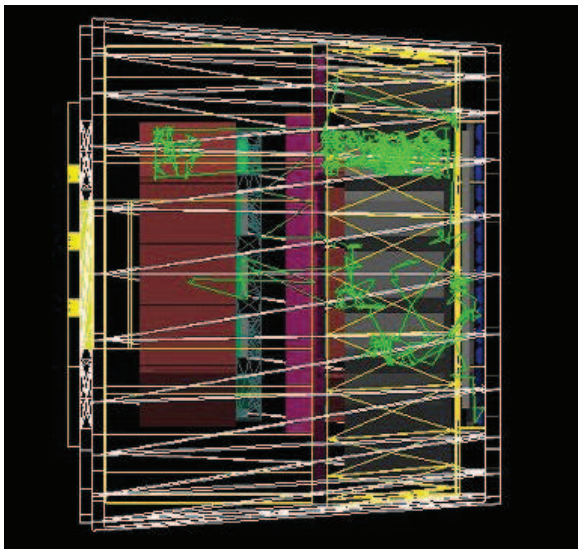}
\caption{ {\scriptsize View of an OpenGL rendering of the geometry used in the LSST camera cryostat Geant4 application, along with an example trajectory of a simulated molecule.  This particular molecule has undergone a far lower than average number of surface interactions and ended up contacting the front surface of a CCD. The geometrical scenario shown is known as the ``baseline'', however the differences between the various scenarios are subtle enough not to dominate such a large-scale view. }}
\label{geo2}
\end{figure}

Three different primary cryostat layouts have been under consideration for LSSTCam.  All three preserve the same basic structure, and differ only in details.  The first, or ``baseline,'' design features RTMs with distinct front and rear electronics boards connected by 2m long cables.  Inside the RTM, the frontmost boards are interspersed with copper thermal planes and connected to the cryoplate. An additional warmer coldplate structure is provided to support and cool the rearmost boards.  A modified design, ``Option 2'' contains the electronics on a single longer board with built-in copper thermal bars, replaces the coldplate with a simpler tubular structure, and in addition makes several other modiﬁcations. ``Option 3'' modifies the design even further by reconfiguring the radiation shrouds and pumping plenum geometry.  To explore the design parameters, further modifications to Option 3 have been examined, for example increasing the diameter of the ion pump ports by either two or four times.  These cases are designated ``Option 3 + 2xI'' and ``Option 3 + 4xI'' respectively.   Figures \ref{geo1} and \ref{geo2} show examples of the detector geometry developed for this application as rendered in the OpenGL viewer.

A number of features within the cryostat have not yet been rendered for the application detector geometry, because they would be too cumbersome to implement and/or are not yet specified, nor are they likely have a significant effect on the bulk molecular transport issues under investigation.  These features include fasteners, cables, thermal straps, fluid lines, hermetic feed-throughs, CCD package kinematic mounts, raft hold downs, and perimeter cutouts in the grid.  Additionally some features have necessarily been simplified to allow for their specification with Geant4 detector construction.  It is unlikely that these simplifications and omissions would have a signiﬁcant effect on the bulk vacuum molecular transport issues being modeled, because they represent a small portion of the total surface area and are mostly perturbations on top of other surfaces.

For this application scoring has been implemented such that each particle interacts with surfaces until it reaches one of three fates --- either contacting the front surface of the CCDs in the focal plane, contacting the surface of a getter pump located near the focal plane, or exiting the cryostat through one of two ion pump ports.  These three fates, any of which ends any given molecule's trajectory, are determined through the use of regions specified in the detector construction as discussed in \S \ref{ao}.  As an indication of the complexity of the LSSTCam geometry, it is found that molecules can have anywhere from less than one hundred to hundreds of thousands of surface encounters before encountering one of these three terminating regions.  

The probabilities for each of these outcomes for several alternate geometries under consideration are presented in Table 2; each is based on simulations of 1000 particles.  The statistical uncertainties on these values can be obtained as outlined in \S \ref{confirm} and are $\sigma$=1\%.  These results show that in the first three geometries most molecules ($>$85\%) eventually reach the focal plane region (i.e. touch the getter pumpus or the front surface of the CCD detectors) rather than exiting through the rear ion pump ports.  Once in that region, they then can encounter either the getter pump or the front surface of the CCD detectors first.  From this one concludes that the getter pumps will need to have a large enough capacity to handle this fraction of the flow. It also suggests these pump surfaces should be made larger or be better positioned.  Alternately, modifications to the geometry to make molecular transport into the ion pumps more likely is warranted.  The simplest modification would be to increase the diameter of the ion pump ports, and as shown in Table II this by itself greatly increases the likelihood of a molecule exiting through an ion pump port before reaching the focal plane region. The ability to readily test the effects of such modifications makes a strong case for this Geant4 modeling and simulation approach.

\section{Discussion} \label{disc}

\begin{figure}
\includegraphics[width=3.0in]{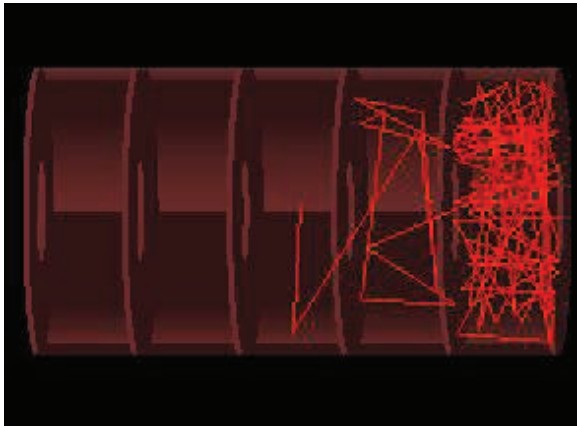}
\caption{{\scriptsize View of an OpenGL rendering of the geometry of a Geant4 application representing a section of accelerator pipe, where regularly spaced irises form cavities, as discussed in \S \ref{disc}.  A trajectory of a molecule created with initial position on the wall of the middle section and with random initial momentum is shown.  }}
\label{beamcavity}
\end{figure}

Geant4 provides a flexible and well-documented framework to simulate particle transport and interactions across complex geometries.  These features, in concert with the techniques described herein, provide a unique method to model molecular transport in complex vacuum environments, where analytical calculations or other techniques are impractical.  Such is the case in many real applications like the LSSTCam cryostat.  As demonstrated in \S \ref{confirm}, these general techniques reproduce results from analytical calculations and other Monte Carlo methods that are available for comparison. 

The techniques we discuss are clearly not unique to a particular vacuum geometry and could be used to model molecular transport in any geometry.  Indeed such modeling can also provide additional information as well.  As an example, Figure \ref{beamcavity} shows a rendering of the geometry of another Geant4 application we have developed, that of an electron accelerator beam pipe with regularly spaced irises that form cavities, with pumps located at either end.   This application has been used to determine mean residence lengths of molecules in a given cavity section, and therefore the partial pressures within, given an outgassing rate for the interior surfaces of the pipe.  To sum the path lengths within a given volume, one simply carries the net length through in an implementation of G4UserTrackInformation and adds to it in an implementation of G4UserSteppingAction, where the step length is added if the step terminates within the volume of interest. 

As discussed in \S \ref{setup}, for the LSSTCam cryostat application, we have so far implemented a simple scoring (i.e.: fraction of particles reaching a focal plane structure versus reaching a pump first).  As already indicated, the efficacy of simple changes to the model such as increasing the diameter of the ion pump ports can be readily evaluated.  In further studies, the effects of adding, resizing, and moving pump ports and getters will be evaluated, as well as adding large area cold activated-charcoal getter pumps which may be placed in the transport regions between the front and the rear of the cryostat, but are not yet included in the current models..  More extensive changes to the geometry and design (e.g.: plenum shape, charcoal getters, molecular flow barriers, etc.) will also be studied. 

Thus far, interest has been in the steady state or ``worst-case'' scenario estimates.  However these applications can be easily extended to include temperature-dependent vacuum sticking coefficients representing the mean duration for a molecular species to be adsorbed on a particular surface.  This time dependency can be included in evaluating, for example, whether a particle will actually contribute to building monolayers of contaminants on the front surface of the CCDs within a given interval of time.  Results from this application are crucial for evaluating the transport, and therefore focal plane contamination consequences, of different layouts and features under consideration.  They will also provide a necessary component of the mapping from outgassing of components (that are measured in a test stand\cite{RSI}) to the resulting effect on the contaminant levels adsorbed on the focal plane, and ultimately to the impact on CCD performance via modeling.\cite{SPIE2}

\begin{acknowledgments}
We wish to thank Dennis Wright, Makoto Asai, and Daniel Brandt for their assistance and expertise in developing Geant4 applications, and Gordon Bowden for useful discussions and critique.  LSST project activities are supported in part by the National Science Foundation through Governing Cooperative Agreement 0809409 managed by the Association of Universities for Research in Astronomy (AURA), and the Department of Energy under contract DE-AC02-76-SFO0515 with the SLAC National Accelerator Laboratory.  Additional LSST funding comes from private donations, grants to universities, and in-kind support from LSSTC Institutional Members.
\end{acknowledgments}


\begin{thebibliography}{0}
\bibitem[1]{LSSTover} Z. Ivezic {\it et al}., LSST: from Science Drivers to Reference Design and Anticipated Data Products, 2008.
(arXiv:0805.2366, 2008)
\bibitem[2]{SPIE1} S. Kahn {\it et al}., Design and Development of the 3.2 Gigapixel Camera for the Large Synoptic Survey Telescope, {\it Proc. SPIE --- Ground-based and Airborne Instrumentation for Astronomy III}, 2010.
\bibitem[3]{RSI} J. Singal, R. Schindler, C. Chang, C. Czodrowski, and P. Kim, A Multi-Chamber System for Analyzing the Outgassing, Deposition, and Associated Optical Degradation Properties of Materials in a Vacuum, {\it Rev. Sci. Instrum.}, {\bf 81}:025101--025108, 2010.
\bibitem[4]{AnsysP} H. Howell, J. Wehrle, and J. Jostlein, Calculation of pressure distribution in vacuum systems using a commercial finite element program, {\it Proceedings of 1991 IEEE particle accelerator conference (PAC)}, pp. 2295--2297, 1991.
\bibitem[5]{Davis} D.H. Davis, Monte Carlo Calculation of Molecular Flow Rates through a Cylindrical Elbow and Pipes of Other Shapes, {\it Journal of Applied Physics}, {\bf 31}:1169--1175, 1960.
\bibitem[6]{Vaktrak} Ziemann V. Vacuum Tracking, {\it Proceedings of 1993 IEEE particle accelerator conference (PAC)}, pp. 3909--3911, 1993.
\bibitem[7]{Molflow} R.Kersevan, \& J-L.Pons, Introduction to MOLFLOW+, {\it J. Vac. Sci. Technol. A}, {\bf 27}:1017--1023, 2009.
\bibitem[8]{Geant1} S. Agostinelli et al., Geant4 --- A Simulation Toolkit, {\it Nuclear Instruments and Methods A}, {\bf 506}:250--303, 2003.
\bibitem[9]{Geant2} J. Allison {\it et al}., Geant4 Developments and Applications, {\it  Int. J. Model. Simul. Sci. Comput.}, {\bf 1}:157--178, 2010.
\bibitem[10]{CDMS} R. Agnese {\it et al}., Dark Matter Search Results Using the Silicon Detectors of CDMS II, {\it PRL}, submitted, 2013. (arXiv:1304.4279)
\bibitem[11]{Geant4DNA} S. Incerti {\it et al}., The Geant4-DNA project, {\it PRL}, submitted, 2013. (arXiv:0910.5684)
\bibitem[12]{PMC} S. Ceyer, D. Gladstone, M. McGonigal, \& M. Schulberg, Molecular Beams: Probes of the Dynamics of Reactions on Surfaces, in Physical Methods of Chemistry: Investigations of Surfaces and Interfaces, eds. B. Rossiter \& R. Baetzold, New York:Wiley, pp. 383--449, 1992.
\bibitem[13]{Rettner89} C. Rettner, E. Schweizer, \& C. Mullins, Desorption and trapping of argon at a 2H–W(100) surface and a test of the applicability of detailed balance to a nonequilibrium system, {\it Journal of Chemical Physics}, {\bf 90}:3800--3814, 1989.
\bibitem[14]{Steinruck84} H. Steinruck, A. Winkler, \& D. Rendulik, Angle-resolved thermal desorption spectra for CO and H2 from Ni(111), Ni(110) and polycrystalline nickel, {\it Journal of Physics C}, {\bf 17}:L311--L316, 1984.
\bibitem[15]{Clausing} P. Clausing, The Flow of Highly Rarefied Gases through Tubes of Arbitrary Length, {\it Ann Physik}, {\bf 12}:961--972, 1932.
\bibitem[16]{SPIE2} K. Gilmore {\it et al}., LSST Camera Instrument Modeling, {\it Proc. SPIE --- Modeling, Systems Engineering, and Project Management for Astronomy V}, 2012.
\end{thebibliography}
\end{document}